\documentclass[twocolumn,pra,aps]{revtex4}
\usepackage{graphicx}

\begin{document}

\title{Quantum interference with photon pairs created in spatially separated
sources}

\author{H. de Riedmatten$^\dag$, I. Marcikic$^\dag$, W. Tittel$^{\dag,\ddag}$, H. Zbinden$^\dag$ and N.
Gisin$^\dag$}
%EndAName

\affiliation{$^\dag$Group of Applied Physics-Optique, University
of Geneva,
Switzerland\\
$^\ddag$Danish Quantum Optics  Center, Institute for Physics and
Astronomy, University of Aarhus, Denmark}

\begin{abstract}
 We report on a quantum interference experiment to probe the coherence between two photons
coming from non degenerate photon pairs at telecom wavelength
created in spatially separated sources. The two photons are mixed
on a beam splitter and we observe a reduction of up to 84$\%$ in
the net coincidence count rate when the photons are made
indistinguishable. This experiment constitutes an important step
towards the realization of quantum teleportation and entanglement
swapping with independent sources.
\end{abstract}

\maketitle

\section{Introduction}

The principle of indistinguishability  is at the heart of the
quantum physical description of the world. It leads to the well
known phenomenon of interference: if two or more processes lead to
indistinguishable detection events, the probability amplitude of
the different processes add coherently and an interference term
appears \cite{feynman,GHZ}. In addition to the most well-known
single photon or first order interference, interference in the
coincidence detection of two or more photons can also be observed.
The so called second-order or two photon interference has been
used for instance to highlight the bosonic nature of photons
\cite{hong87} and to demonstrate non-local effects between photons
forming entangled pairs. Moreover it is as the origin of the new
field of quantum information processing (for a recent overview
concerning the last two points, see e.g. \cite{tittel01}).

Observing second order quantum interferences with photons without
common history is a very important issue, since this forms the
basis of entangling those photons through a so called
interferometric Bell state measurement
\cite{michler96,zukowski95}. The simplest way to observe such
interference is to mix the independent, however indistinguishable
photons on a beam splitter (i.e one photon per input mode). In
this case, the probability amplitudes of both photons being
transmitted or both reflected cancel each other and the two
photons will always be detected in the same output mode. This is
valid of course only if the photons coming from the two sources
become indistinguishable after the beam splitter, i.e. if they are
described by identical polarization, spatial, temporal and
spectral modes. In other words, when all indistinguishability
criteria are met, the count rate for coincidence detection of two
photons in different output modes of the beam splitter drops to
zero. The first experiments showing this effect were made by
Mandel and coworkers in the end of the eighties \cite{hong87}. The
drop in the coincidence count rate when varying the temporal
overlap between the two photons is often referred to as a "Mandel
dip". In those early experiments, the two photons belonged to one
photon pair generated by parametric down-conversion (PDC). In this
case, the temporal indistinguishability is ensured by the fact
that the two photons are created simultaneously. In cases where
two independent photons (i.e created in different sources or
different PDC events) have to interfere, there are two
possibilities to restore the temporal indistinguishability (i.e.
to ensure that only photons detected in coincidence within their
coherence time contribute). The first and most common way is to
create the photon pairs using ultra-short pump pulses such that
the downconverted photon's coherence time (given by the phase
matching conditions) is superior to the duration of the pump pulse
\cite{zukowski95}. The second possibility is to increase the
coherence time of the photons such that it becomes larger than the
temporal resolution of the detectors. In this way, one can select
the photons arriving at the same time at the beam splitter
directly by their arrival times at the detectors. This method
requires coherence times of the order of a few hundreds of
picosecond, with current avalanche photodiodes detectors
\cite{stucki01}. This can be achieved for instance by using a
sub-threshold OPO configuration \cite{ou99}.
\\
In the context of observing quantum interference between
independent photons, Rarity et al have performed an experiment
where a one photon state obtained by PDC and a weak coherent state
were mixed at a beam splitter \cite{rarity97}. More recently,
experiments where two photons from different pairs interfere at a
beam splitter have been carried out in order to implement quantum
teleportation \cite{bouwmeester97}, entanglement swapping
\cite{pan98,jennewein02}, and to create GHZ states
\cite{bouwmeester99,pan01}. In these experiments, however, the two
photon pairs were created in the same crystal by means of two
subsequent passages of a pump pulse.\\
In this article, we go a step further and report on the
observation of quantum interference with photons from different
pairs created in two spatially separated sources. We use non
degenerate photon pairs at telecom wavelength. This is an
important extension with respect to the previous experiments since
some quantum communication protocols (for instance the quantum
repeater \cite{briegel98}) rely on the use of photon pairs created
at different locations, hence photon pairs from different sources.

\section{Mandel dip with independent PDC sources}

As shown in Fig. \ref{setup}, we create pairs of non degenerate
photons in two non-linear crystals using short pump pulses. The
photons belonging to a pair are separated and two photons from
different sources are superposed on a 50-50 beam splitter. We
label $a^{\dagger}$ and $b^{\dagger}$ ($c^{\dagger}$ and
$d^{\dagger}$) the creation operators of the two input (output)
modes respectively. Unitarity implies that the phase difference
between a reflected and a transmitted photon is $\frac{\pi}{2}$.
For a 50-50 beam splitter, the evolution is thus
$a^{\dagger}\rightarrow
\frac{1}{\sqrt{2}}(c^{\dagger}+id^{\dagger})$ and
$b^{\dagger}\rightarrow
\frac{1}{\sqrt{2}}(ic^{\dagger}+d^{\dagger})$.

Suppose that we have one photon in each input mode, i.e the
following Fock state:
\begin{equation}
\left| \psi _{in}\right\rangle
=a^{\dagger}b^{\dagger}\left|0\right\rangle
\end{equation}
 After the beam
splitter the state becomes:
\begin{eqnarray}
 \label{bs}\left| \psi _{out}\right\rangle =\frac{1}{2}(i
(c^{\dagger})^{2}+i
(d^{\dagger})^{2}+c^{\dagger}d^{\dagger}-c^{\dagger}d^{\dagger})\left|
0\right\rangle\\\nonumber
 =
 i\frac{1}{\sqrt{2}}\left|2\right\rangle_{c}\left|0\right\rangle_{d}+i\frac{1}{\sqrt{2}}\left|0\right\rangle_{c}\left|2\right\rangle_{d}\\\nonumber
\end{eqnarray}
We thus find that for a 50-50 beam splitter, the two probability
amplitudes corresponding to both photons transmitted and both
photons reflected -- i.e to both photons in different output ports
-- cancel out and the coincidence rate drops to zero. This
description is valid of course only if the photons become
completely indistinguishable after the beam splitter. If we delay
the photon from one source with respect to the other one, we loose
temporal indistinguishability, and the destructive interference
diminishes. We define the visibility of the Mandel dip as follows
:
\begin{equation}
V_{dip}=\frac{I_{\max }-I_{\min }}{I_{\max }}
\label{vis}
\end{equation}
In the case of two PDC sources, there are different possibilities
to create 2 photon pairs at the same time: either one creates one
pair in each source, or two pairs in one source and none in the
other one. As already said, in order to ensure temporal
indistinguishability, the coherence time of the down-converted
photons must be larger than the duration of the pump pulses. This
implies that the pairs created within the same laser pulse and
crystal are subject to stimulated emission \cite{Lamas01}. The
output state of a non degenerate PDC follows the distribution
\cite{walls94,scarani02}:
\begin{equation}
\left| \Psi\right\rangle= e^{-g} e^{\Gamma
A^{\dagger}B^{\dagger}}\left|0\right\rangle=\sum_{n} \frac{(tanh
\zeta)^{n}}{cosh \zeta}\left|n_{A},n_{B}\right\rangle \label{dist}
\end{equation}
where $A^{\dagger}$ and $B^{\dagger}$ are the creation operators
for the PDC modes A and B, and  $\left|n_{A},n_{B}\right\rangle$
corresponds to n photons in PDC modes A and B respectively. The
parameter $\zeta$ is proportional to the amplitude of the pump
field and to the non-linear susceptibility $\chi^{(2)}$, $g =
ln(cosh \zeta)$ and $\Gamma = tanh\zeta$. In the limit of small
$\zeta$, we have:
\begin{equation}
\left| \Psi\right\rangle=(1-\frac{\zeta^{2}}{2})\left|
0\right\rangle+\zeta\left| 1,1\right\rangle+\zeta^{2}\left|
2,2\right\rangle+O(\zeta^{3}) \label{dist1}
\end{equation}
If $P_{i}(I)=|\zeta|^{2}$ is the probability of creating one pair
per pulse with a pump intensity I in source i, then the
probability of creating 4 photons per pulse in source i by
stimulated emission is $|\zeta|^{4}=P_{i}^{2}(I)$ . The
probability of simultaneously creating one pair in each crystal is
$P_{1}(I)P_{2}(I)$. Assuming that $P_{1}(I)=P_{2}(I)$ the
4-photons state can then be written as follows (not normalized):
\begin{eqnarray}
\nonumber\left| \Psi _{4ph}\right\rangle =\left|
2_{1},0_{2}\right\rangle
_{A}\left| 2_{1},0_{2}\right\rangle _{B}\\
\nonumber+\left|1_{1},1_{2}\right\rangle _{A}\left|
1_{1},1_{2}\right\rangle _{B}\\
 +\left|
0_{1},2_{2}\right\rangle _{A}\left| 0_{1},2_{2}\right\rangle _{B}
\end{eqnarray}
where here for instance $\left| n_{_{1}},m_{_{1}}\right\rangle
_{A}$ means that we have n photons in source 1 and m photons in
source 2 created in the PDC mode A. It is important to notice
that, due to stimulated emission, the amplitudes of each of the 3
terms are the same. This means that, the probability of creating 4
photons per pulse in a specific source is the same as the
probability of creating simultaneously 2 photons in source 1 and 2
photons in source 2.

If there is no interference, the photons arriving at the beam
splitter will split in half of the cases. The probability of
detecting a coincidence outside the dip (od) is thus proportional
to:
\begin{equation}
P_{od}\propto \left(
\frac{P_{1}^{2}}{2}+\frac{P_{2}^{2}}{2}+\frac{P_{1}P_{2}}{2}\right)
=\frac{3P^{2}}{2}
\end{equation}
where $P_{1}=P_{2}=P$ is the probability of creating one pair per
pulse per source. The first two terms represent the creation of
four photon in either source, the last term the creation of one
pair per source. Inside the dip (id), the contribution of the
events where one pair per source is created drops to zero (Eq.
\ref{bs}). We thus have:
\begin{equation}
P_{id}\propto \left(
\frac{P_{1}^{2}}{2}+\frac{P_{2}^{2}}{2}+0\right)=P^{2}
\end{equation}
Finally, the visibility is :
\begin{equation}
V=\frac{P_{od}-P_{id}}{P_{od}}=\frac{3P^{2}-2P^{2}}{3P^{2}}=\frac{1}{3}
\end{equation}
\noindent The maximum theoretical visibility is thus V=33\%,
because, in this case, we cannot discard the events where both
photon pairs are created in the same crystal. Note that this
demonstrate 2 photon interference between 2 thermal sources
\cite{mandel83,kuo91,ou99b}. However if we detect the two
remaining photons as well (4-photon coincidences), we post-select
only the events where we create one photon pair per crystal.
Therefore the maximum theoretical visibility is V=100 \%. A
detailed theoretical analysis can be found in \cite{ou99b}.
Obviously, this is valid only if we can neglect the probability of
creating three pairs at the same time, two in one source and one
in the other one. We will discuss this case later.

\section{Experimental setup}
\begin{figure}[h]
\includegraphics[width=8.43cm]{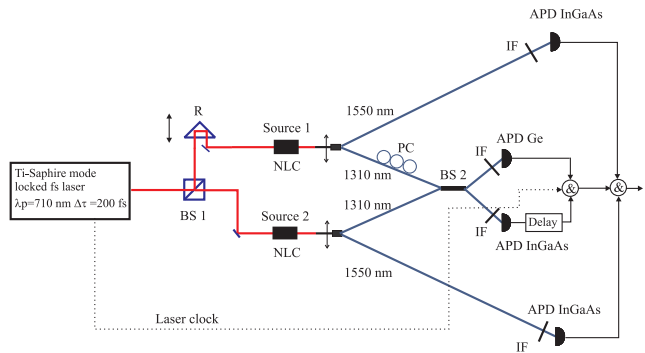}
\caption{Experimental setup used to observe quantum interferences
with photons coming from different sources} \label{setup}
\end{figure}

The experimental setup is shown in figure \ref{setup}. The pump
laser is a Ti-Sapphire mode-locked laser (Coherent Mira),
operating at a wavelength of 710 nm and generating 150
femto-second pulses with 4.5 nm bandwidth (FWHM). The pump beam is
split by a beam splitter (BS 1), and the two output modes are used
to pump two 10 mm Lithium Triborate(LBO) non linear crystals. The
average pump power is $\overline{P}_{pump}\approx 40$ mW per
crystal. In each crystal, non-degenerate collinear photons pairs
at telecom wavelength (1310 nm/1550 nm) are produced by type I
parametric down conversion. The photons are then coupled into a
standard optical fiber and separated using a wavelength division
multiplexer (WDM). The 1310 nm photons are directed to a 50/50
fiber coupler (BS 2). The length of optical fibers before the
beam-splitter is equalized within a few hundreds of $\mu m$, in
order to have the same chromatic dispersion in the two input
modes. To ensure equal polarization for the photons coming from
either source, we use a fiber optical polarization controller (PC)
inserted in one arm. In order to insure temporal
indistinguishability, the optical distance between BS1 and BS2
must be the same within the coherence length of the down converted
photons. To vary this distance and hence to vary the temporal
overlap, we use the retroreflector R that is mounted on a
micrometric translation stage.

The photons are detected with photon counters. One output of BS 2
is connected to a passively quenched Germanium avalanche
photodiode (Ge APD) cooled with liquid nitrogen. The quantum
efficiency of The Ge APD is 10$\%$ for 40 kHz dark counts. The
dark counts are reduced to around 3 kHz by making a coincidence
with a 1 ns clock signal delivered simultaneously with each laser
pulse ($t_{0}$). The signal count rate on the Ge APD is 40 kHz.
The other output is connected to a Peltier cooled (T=220K) Indium
Gallium Arsenide (InGaAs) APD, operating in so-called gated mode
\cite{stucki01}. This means that it is only actived within a short
time window (100 ns) after a Ge-$t_{0}$ coincidence. InGaAs APDs
feature a quantum efficiency of around 30$\%$ for a dark count
probability of $\approx 10^{-4}$ per ns. Interference filters (IF)
(10 nm FWHM centered at 1310 nm) are placed in front of the
detectors to increase the coherence length (time) of the down
converted photons to 75 $\mu m$ (250 fs). Using the side peaks
method developed in \cite{marcikic02}, we measure the probability
to create one photon pair per pulse in the spectral range given by
the filters to be of around 4$\%$. The signals from the APD's are
finally sent to detection and fast (1 ns) coincidence electronics.
A coincidence between the two detectors for 1310 nm photons and
the laser clock ($t_{0}$)is referred as a 3-fold coincidence.

In order to obtain a Mandel dip with 100\% visibility, a 4-photon
coincidence using also the two others photons at 1550 nm is
necessary, in order to post-select only the interfering events.
The photons at 1550 nm are detected with InGaAs APDs, gated using
a 3-fold coincidence (2 photons at 1310 nm + $t_{0}$). We thus
speak of 5-fold coincidence in this case. The very low gate rate
imposed by this scheme allows us to avoid problems with
afterpulses of the InGaAs APDs \cite{stucki01}. Interference
filters (10 nm FWHM centered at 1550 nm) are  also placed in front
of the 1550 nm detectors, in order to reduce the probability of
detecting events where 3 pairs are created simultaneously.

\section{Results}

\begin{figure}[h]
\includegraphics[angle=0,width=8.43cm]{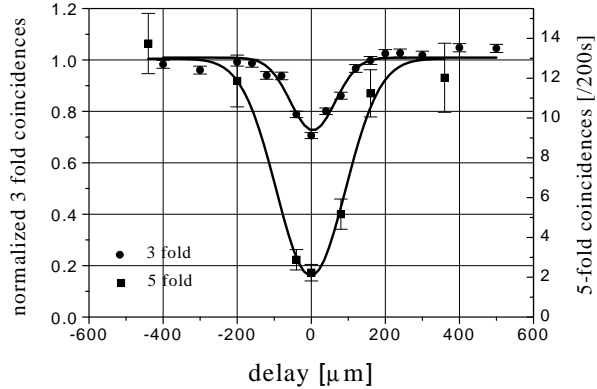}
\caption{Coincidences count rate as a function of the delay of one
photon. The circles are 3-fold coincidences (2 photons + laser
clock) and the squares are 5-fold coincidences (4-photon + laser
clock). The net visibility of the dip is $28 \%$ and 84$\%$
respectively. The integration time for 3-fold coincidence is 200
s, while for 5-fold coincidence, it varied from 30 to 60 min, such
that the statistical error on the counts is around 10$\%$. The
5-fold rate is scaled to 200 s.} \label{dip5fold}
\end{figure}

Figure \ref{dip5fold} shows the coincidence count rate as a
function of the position of the retroreflector R, i.e. of the
delay of one photon. The circles represent the 3-fold coincidences
(2 photons at 1310 nm+ $t_{0}$) and the squares the 5-fold
coincidences (4 photons +$t_{0}$). We measure around 160 net
3-fold coincidences and around 0.06 net 5-fold coincidences per
second outside the dip. Accidental coincidences (around 20 3-fold
and 0.015 5-fold coincidences per second) are already subtracted
in the presented data. 3 fold coincidences slightly vary ($\approx
10 \%$), probably due to temperature variation in the lab during
day time measurement, and are normalized with the square of single
count rate of the Ge APD. The 5-fold curve has been measured
during the night, when the temperature was more stable. Thus,
count rates variations were smaller and raw data can be used
without
normalization. \\
The shape of the dip is given by the convolution of the two
wave-packets arriving at the beam splitter \cite{hong87}. The
spectral transmission of the IF has been measured to be well
 approximated by a gaussian. We thus fitted our data with
the following function :
\begin{equation}
R_{c}(\tau )=S\left( 1-Ve^{-\frac{\tau ^{2}}{2\sigma _{\tau }^{2}}}\right)
\end{equation}
where $S$ is the number of coincidences outside the dip, V the visibility, $%
\tau $ the optical delay and $\sigma _{\tau }$ the $1/\sqrt{e} $
half width of the gaussian function. Due to the convolution
product, the expected FWHM of the dip is $\sqrt{2}  l_{c}$, where
$l_{c}$ is the FWHM coherence length of the down converted
photons, given by the IF.

We obtain a raw (i.e without subtracting accidental coincidence)
visibility of $V= (21 \pm 1)\%$ for the 3-fold coincidence. When
subtracting accidental coincidence, it increases to $V=(28 \pm
2)\%$, which is close to the theoretical visibility of $33\%$. The
FWHM of the gaussian fit is of $142 \pm 15 \mu m $. This is
slightly larger than the expected value ($\sqrt{2}l_{c} = 107 \mu
m$ for 10nm IF at 1310 nm). When fitting the 5-fold coincidence
curve, we obtain a raw visibility of $(77\pm 2.5)\%$, which
increases to $(84\pm 2.5)\%$ when subtracting accidental
coincidences. The FWHM of the gaussian fit ($222\pm25\mu m$) is
larger than for the 3-fold curve. This can be qualitatively
understood by the fact that the 10 nm IF at 1550 nm reduces the
bandwidth of the 1310 twin photons to $\approx 7$ nm by energy
conservation. \\
Various reasons could explain the difference between the
theoretical and experimental visibilities. The main reason is the
probability of detecting events where 3 pairs are created
simultaneously, 2 in one source and 1 in the other source. A
calculation similar to eq (\ref{bs}), starting from an input state
$\left| \psi _{in}\right\rangle =
\frac{1}{\sqrt{2}}(a^{\dagger})^2b^{\dagger}\left| 0\right\rangle$
shows that these events will indeed induce spurious coincidences
and thus reduce the visibility of the dip. To estimate this
maximal visibility, we calculate the maximal and minimal
coincidence count rates ($I_{max}$ and $I_{min}$) for all cases
leading to a 5-fold coincidence, and insert them into eq.
(\ref{vis}) with the corresponding probabilities, computed from eq
(\ref{dist}). We neglect the events where more than 3 pairs are
created simultaneously. The finite quantum efficiency $\eta$ of
detectors is taken into account, in the sense that the probability
of having a click when two photons arrive at the detector is
$2\eta-\eta^{2} \approx 2\eta$ for small $\eta$. In the case where
the transmissions of the two inputs modes of the beam splitter are
the same, a simple but lengthy calculation leads to :
\begin{equation}
V_{max}=\frac{1+8P}{1+12P}
\end{equation}
where P is the probability of creating one pair per pulse. For
$P=4\%$, we find $V_{max}=89\%$. \\Furthermore, even in the events
where we have one photon per input mode, the following reasons,
listed by order of importance, might induce distinguishability
between the photons, and thus diminish the visibility of the dip .
There might be remaining temporal distinguishability due to
relatively large (10 nm) filtering of the downconverted photons
(compared to the pump bandwidth)\cite{ou99b}. Moreover, a slight
difference in the polarization of the two photons when arriving at
the beam splitter could result in a which-path information.
Finally, different phase-matching conditions in the two crystals
could result in photon pairs with different spectra. Those
difference might not be completely cancelled with
the 10 nm interference filters.\\

\section{Conclusion}
We observed quantum interference with photon pairs at
telecommunication wavelengths created by parametric down
conversion in spatially separated sources. Two photons, one from
each source were mixed on a beam splitter. When recording 2 photon
coincidences and varying the temporal overlap between the two
photons, we observed a Mandel-type dip with visibility of $(28\pm
2)\%$.  This is close to the maximum visibility of $33\%$, limited
by the impossibility to discard the events where 2 pairs are
created in the same crystal. Recording 4 photon coincidences and
thus post-selecting only events where at least one pair is created
in each source, we obtained a net visibility of $(84\pm2.5)\%$,
close to the theoretical value. This experiment constitutes a
first step towards the realization of quantum teleportation and
entanglement swapping with independent sources. However, note that
truly independent sources require the use of independent but
synchronized fs laser. Although this is nowadays commercially
available  \cite{menlo02}, synchronization of two fs laser at
large distance still has to be demonstrated.
\\
\emph{Note} It was recently brought to our attention that a
similar experiment was reported in the conference QELS 99 by Rhee
and Wang \cite{rhee99}

\subsection*{Acknowledgements}
The authors would like to thank Claudio Barreiro and Jean-Daniel
Gautier for technical support. Financial support by the Swiss NCCR
Quantum Photonics, and by the European project QuComm is
acknowledged. W.T. acknowledges funding by the ESF Programme
Quantum Information Theory and Quantum Computation (QIT).

\end{document}